# DYNAMIC PERCOLATION OF ELECTRIC CONDUCTIVITY AND A TREND TOWARDS FRACTAL SKELETAL STRUCTURING IN A RANDOM ENSEMBLE OF MAGNETIZED NANODUST

A.B. Kukushkin,[1] N.L. Marusov,[1] V.S. Neverov,[1,2] I.B. Semenov,[1]
K.V. Cherepanov,[1] P.V. Minashin[1,2]

[1]NFI RRC "Kurchatov Institute", Moscow 123182, Russia
[2]Moscow Engineering Physics Institute, Moscow 115409, Russia

**Abstract.** Numerical modeling of electrodynamic aggregation is carried out for a random ensemble of magnetized nanodust taken as a many body system of strongly magnetized thin rods (i.e., one-dimensional static magnetic dipoles), which possess electric conductivity and static electric charge, screened with its own static plasma sheath. The self-assembling of quasi-linear filaments from an ensemble of randomly situated basic blocks and the electric short-circuiting between biased electrodes are shown to be supported by the alignment of blocks in an external magnetic field. Statistical analysis of short-circuiting time allows tracing the dynamic percolation of electric conductivity and shows a decrease of percolation threshold for volume fraction (vol%), as compared with the observed percolation of carbon nanotubes (CNTs) in liquids and polymer composites. Modeling of short-circuiting stage of evolution is continued with tracing the dynamics of pinching of electric current filaments to show the interplay of all the magnetic and electric mechanisms of filaments' networking. A trend towards a fractal skeletal structuring (namely, repeat of original basic block at a larger length scale) is illustrated with the evidence for generation of a bigger magnetic dipole.

## 1. Introduction

Here we study the possibility of self-assembling of fractal filamentary structures from a magnetized electroconductive nanodust. We use the 3-D numerical model [1] for a many body system of strongly magnetized thin rods (i.e. 1D static magnetic dipoles). Each block possesses the longitudinal electric conductivity and the electric charge, statically screened with its own plasma sheath. Numerical modeling of $\sim 10^2$-$10^3$ such dipoles has shown [2-5] the possibility of electrodynamic self-assembling of a tubular skeletal structure from an ensemble of initially-linear artificially-composed filaments, linked to the biased electrodes. To substantiate such initial conditions, the possibility of self-assembling of quasi-linear filaments (and closing of the electric circuit) was studied [6] for an initially random ensemble of basic blocks in the magnetic field of a plasma filament with internal longitudinal magnetic field (i.e., nearly force-free internal magnetic configuration).

This research is aimed at testing the capability of the model [1] to describe self-assembling of a fractal skeletal tubular structure. Such a process was suggested [7-9] to be responsible for the unexpected longevity of straight filaments, and their networks, revealed [10] in the Z-pinch gaseous electric discharges. This hypothesis predicted the macroscopic fractal structures with basic topological building block of tubular form (with presumably carbon nanotube (CNT) at the nanometer length scales), which is successively self-repeated at various length scales (see summary of experimental evidences [11] for self-similarity of skeletal structures in the range $10^{-5}$cm-$10^{23}$cm, that included the high-current electric discharges in various laboratory experiments, severe weather phenomena and space; see also the surveys [12,13] and web pages [14,15]).

In the model [1], magnetic aggregation of nanoparticles in electric discharges is treated under assumption [8,9] of the possibility of nanotubular structures to acquire static magnetic dipole moment (presumably, at the initial stage of discharge, e.g., during electric breakdown, or in a similar transient process). Note that the magnetism of basic blocks may not be long-lived like that in the case of ferromagnetism (or similar phenomena of unexpected magnetism of carbon, studied experimentally and theoretically -- for summary see, e.g., monograph [16]): it would be sufficient to trap magnetic flux for a period of welding of (i.e. formation of covalent bonds between) the tips of basic blocks during electrodynamic aggregation. Despite the existing experimental evidences and theoretical models for a strong magnetism of nanoblocks (see some references in [11,12]) need much stronger tests and confirmations, they justified explicit demonstration of the capability of magnetized nanotubular blocks to self-assemble the tubules of higher generations [8,9] (see also [17]) and sustain the integrity of the assembled skeleton. The integrity of a hypothetical tubular skeleton (i.e. the tubule of the 2-nd generation, in terms of the approach [8,9]), virtually-assembled from the above-mentioned basic blocks, was shown in the frame of the model [1].

Further substantial development of the hypothesis [7-9] for a fractal macroscopic skeleton, which repeats the CNT structure at larger length scales, may be found in the papers [18-31] where mechanical and electrophysical properties of this hypothetical, virtually-assembled nanomaterial, named in [18] as "super carbon nanotube", have been studied theoretically with various numerical methods.

In this paper we continue studying theoretically the ways to *fabricate* a wide class of fractal skeletal nanomaterial (not as ideal fractal as super CNT but still a fractal) via *electrodynamic aggregation* of the above mentioned basic blocks with a strong contribution of internal self-organization processes (*self-assembling*).

## 2. Dynamic percolation of electric conductivity in a random ensemble of a magnetized nanodust

First, we analyze the problem of filament self-assembling in the presence of a (quasi)homogeneous external magnetic field. The process under consideration models the initial stage of skeletal structure formation in the laboratory electric discharges in the presence of external magnetic field (in particular, in toroidal magnetic systems).

Numerical modeling of evolution of a random ensemble of basic blocks with above-mentioned properties is focused at the influence of quasi-homogenous background magnetic field on self-assembling of filaments from randomly situated dust particles, including a comparison with the case of self-assembling in the presence of magnetic field of a plasma electric current filament [6]. The closing of the electric circuit ("short-circuiting") is defined as the emergence of, at least, a single filament whose ends lie closer than $L/2$, where $L$ is basic block's length, to the edge z-planes of the initial box (see Fig 1, left). The filament is defined as a chain of blocks whose closest tips permanently lie within the distance $D$ (where $D=0.06 L$ is diameter of the block), which describes the characteristic radius of the potential well of the model pair interaction of the tips (compilation of magnetic attraction and elastic repulsion of the tips, see Fig. 1 in [1]). The results of our analysis are illustrated with Figs. 1-4, where time and magnetic field are expressed, respectively, in the units of $t_0$ and $B_0$:

$$t_0 = \frac{\sqrt{mL^3}}{Z_{M0}e} \sim \left(\frac{L}{10\,nm}\right)^2 \sqrt{\frac{D_{CNT}}{1\,nm}} \frac{1}{Z_{M0}} (n\,\text{s}), \qquad (1)$$

$$B_0 = \frac{Z_{M0} e}{L^2} \sim Z_{M0} \left(\frac{10\,nm}{L}\right)^2 5 \cdot 10^{-2}\ (T) \tag{2}$$

where $2m$ and $L$ are the mass and the length of each basic block; $Z_M = \Phi/4\pi e$ is the effective magnetic charge (in the units of electron charge $e$) on the tip of 1-D magnetic dipole; $\Phi$ is magnetic flux, trapped in the 1-D dipole (below we take $Z_{M0} = 1$), that is close to the average effective magnetic charge. In Eq.(1), for $t_0$ we give also its numerical estimate for particular case of single-walled carbon nanotube of diameter $D_{CNT}$.

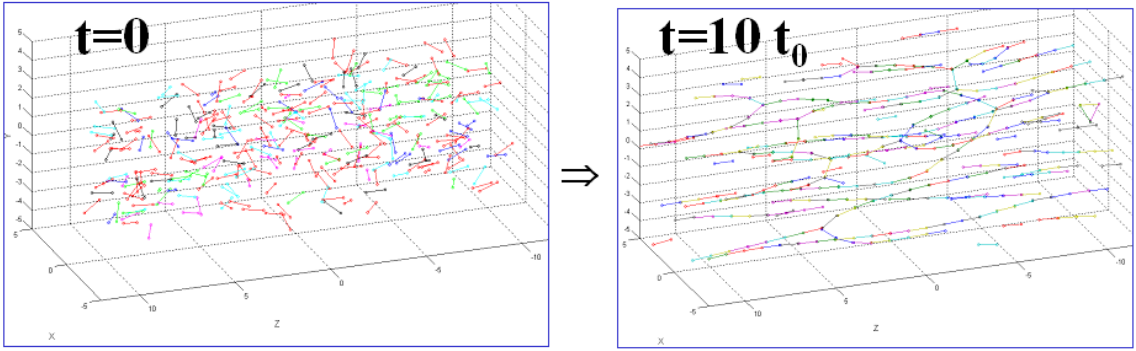

**Fig. 1.** Initial position (left) and that at time $t \sim 10\,t_0$ (right) for the ensemble of 300 blocks, randomly situated at $t=0$ in the box $12 \times 12 \times 20$ (in the units of dipole's length $L$) in external magnetic field $B=B_z=3B_0$. Magnetic charge $Z_M$ is random in the interval $\{0.67, 2\}$ (with average value $<Z_M> = 4/3$, respectively), electric charges $Z_{ei}=Z_{Mi}$ ($i=1\div300$), electrical screening radius $r_D=L$, brake coefficients for tip's collision, $(k_{br})_{dd} = 100$, and for brake in the ambient medium, $(k_{br})_{dm} = 3$ (see [1] for notations).

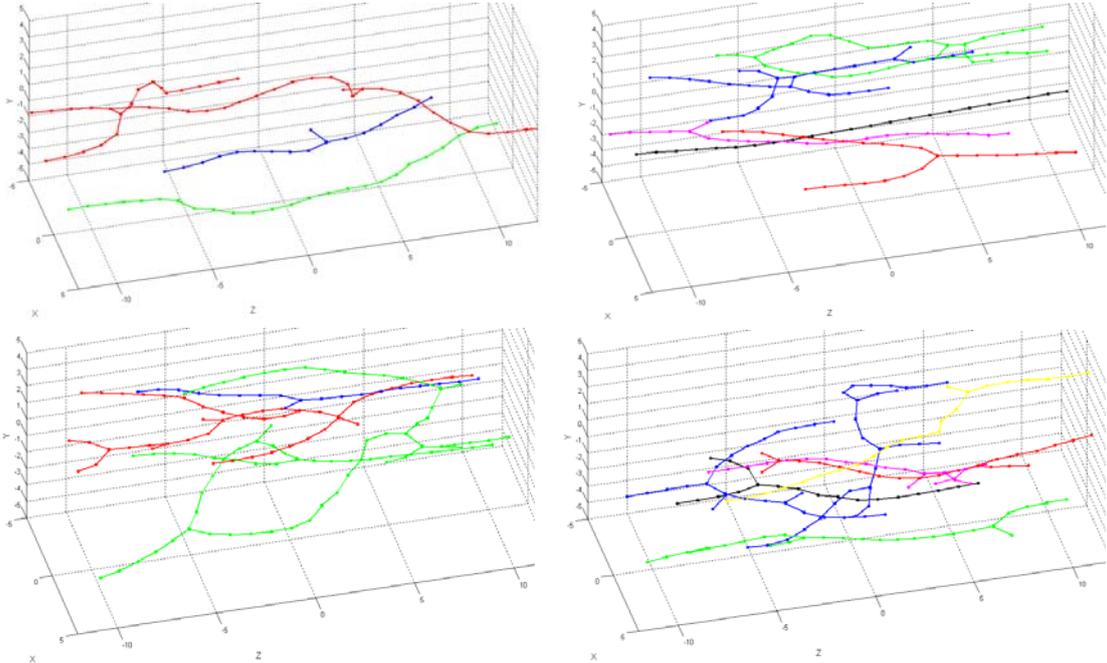

**Fig. 2.** Four typical examples of the *longest* filaments at time $t \sim 10\,t_0$ in the ensembles with different random initial positions at $t=0$. Other parameters are similar to those for Fig 1.

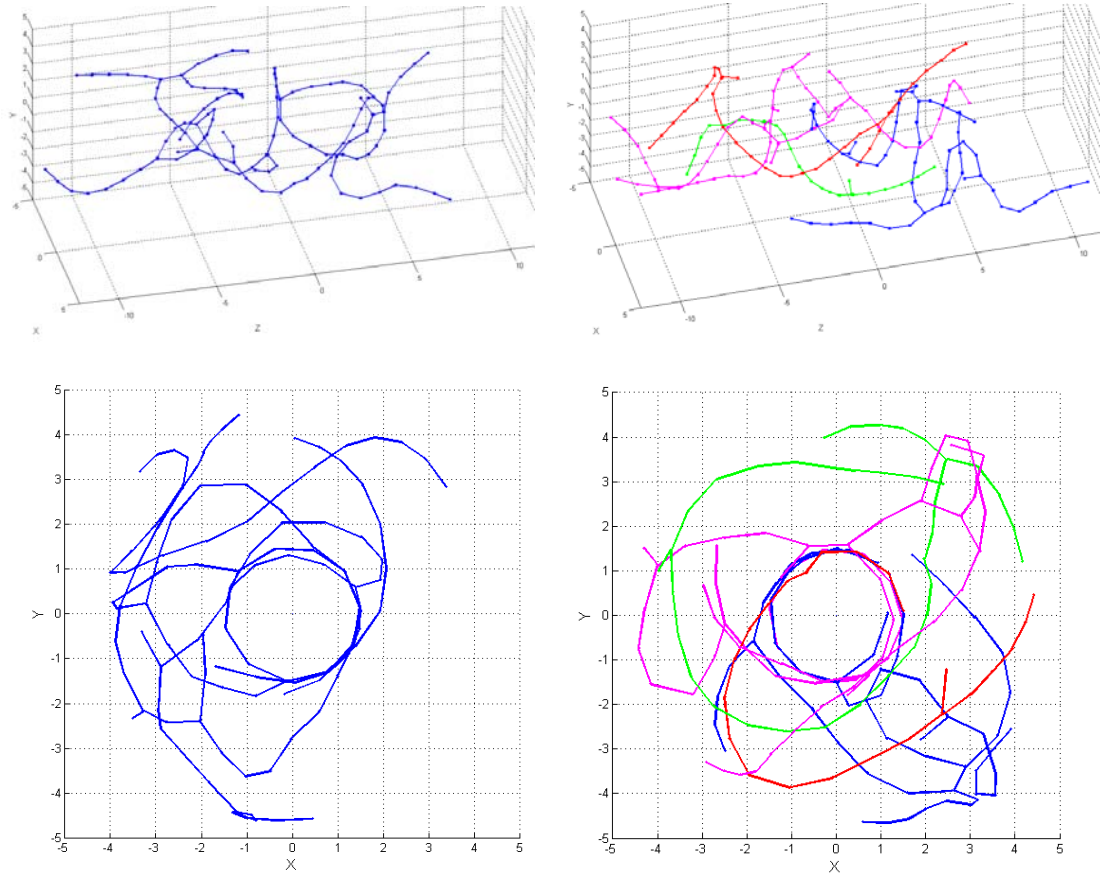

**Fig. 3.** The same as in Fig. 2 but for $B_z=B_0$ and in the presence of magnetic field of a plasma electric current filament (z-directed, with center at x=y=0) with a homogeneous distribution within cylinder of radius $R_{pl.}= 1.5L$ and total longitudinal electric current $J_{zPlas}=1.5cZ_{M0}e/L$. The cases of *random* distribution of magnetic charges (left) and *quantized* distribution, $Z_M=1$ and $Z_M=2$ with ratio 2/1, (right) are presented (here the average value is the same, $\langle Z_M \rangle = 4/3$). Lower figures are the top views of upper figures.

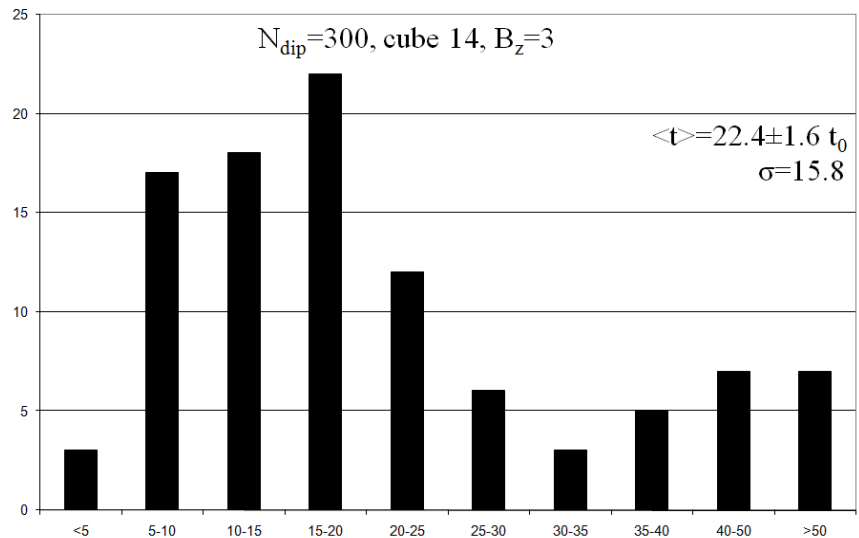

(1)

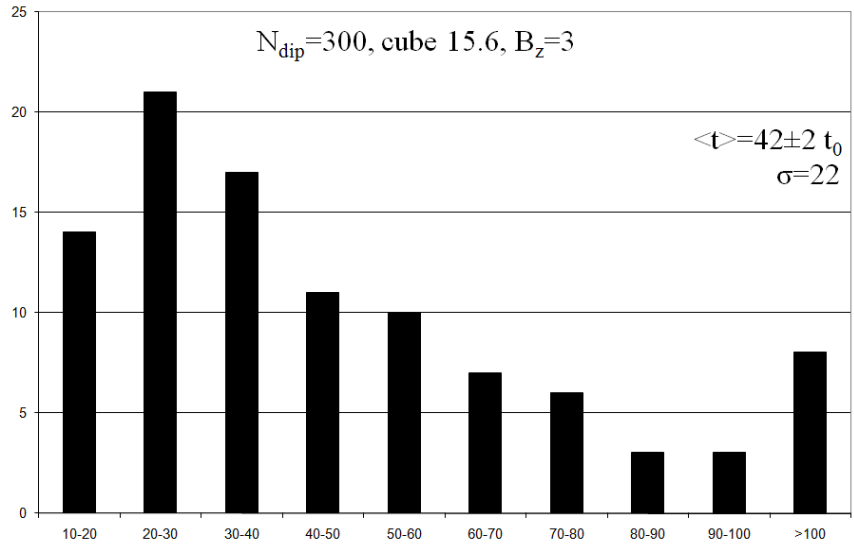

(2)

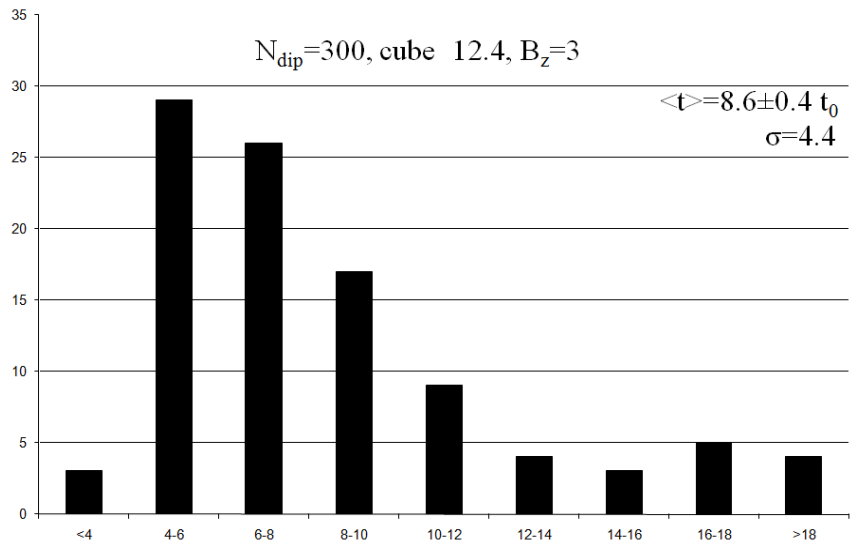

(3)

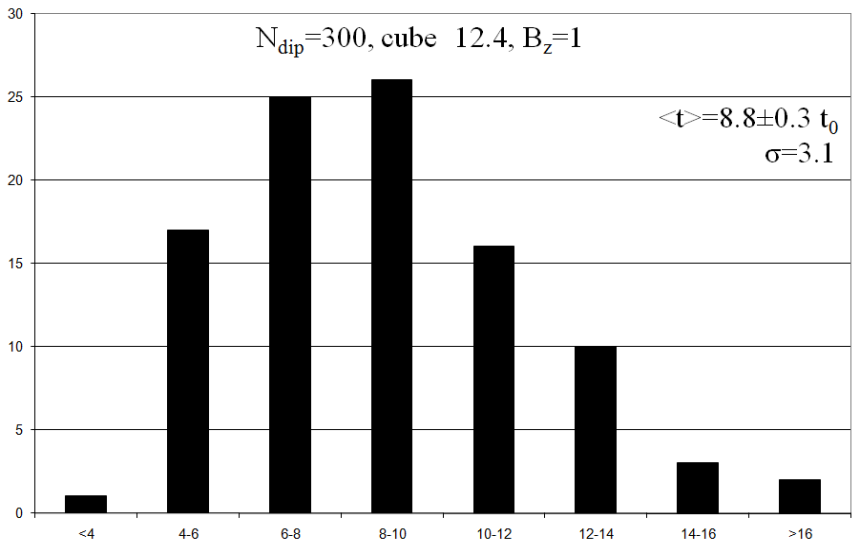

(4)

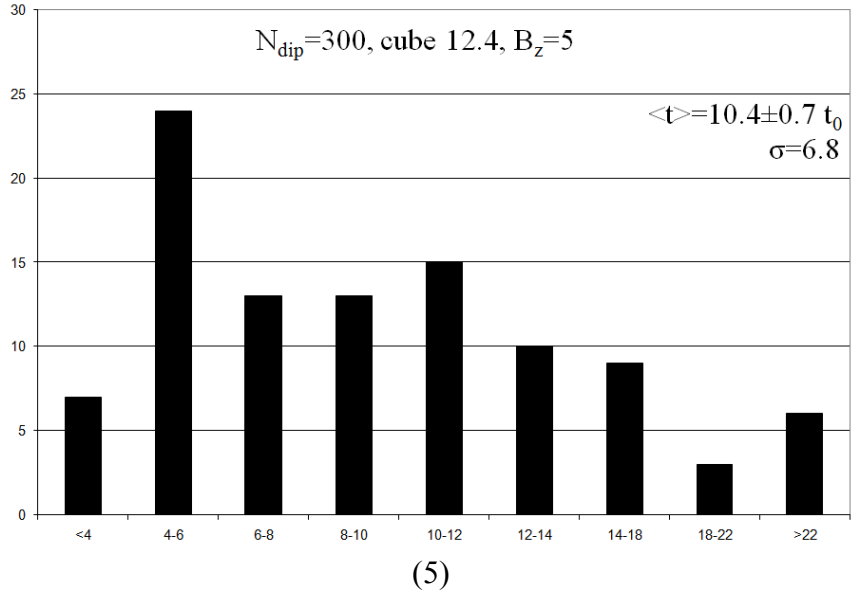

(5)

**Fig. 4.** Statistics of the time needed for closing the electric circuit ("short-circuiting") with at least a single filament for conditions of Fig. 1, but another geometry of initial box (a cube of the size 12.4, 14, and 15.6). For the size 12.4, a scan over $B_z/B_0 = 1, 3, 5$ is presented. Here $<t>$ is the average time of short-circuiting, with confidence probability 0.7, σ is the root-mean-square deviation. Everywhere, 100 computations were done up to time 100 $t_0$. Local maxima in the pockets «35-40» and «16-18» in the histograms 1,3 illustrate the scale of statistical errors.

Figure 4 illustrates how the alignment of the magnetized blocks along a static external magnetic field works for short-circuiting: it suppresses the trend towards isotropic networking of self-assembled filaments and thus shortens the way towards short-circuiting. This trend, obviously expected if comparing with the case of no external magnetic field, is illustrated here with the impact of azimuthal magnetic field of a plasma electric current filament, which may complicate the short-circuiting (cf. Fig. 2 vs. Fig. 3) or, with increasing plasma current, even prevent it.

The timescale of short-circuiting substantially depends on the initial density of blocks, $n_{dip}^{(0)}$ (cf. Fig. 4, histograms 1-3). For $n_{dip}^{(0)} \approx 0.16\ L^{-3}$ (histograms 3-5), it is comparable with that for subsequent stages of self-assembling (namely, $t \sim 10\ t_0$), which have been modeled in [3,4] with simulating the dynamics of a bunch of initially-linear filaments linked to a biased electrodes. However, for twice lower $n_{dip}^{(0)}$ this time increases by a factor of 4-5 (histograms 1,2).

Dependence of short-circuiting time on the strength of external magnetic field, $B_z$, appears to be weaker (cf. histograms 3-5 in Fig. 4). Nevertheless, it is possible to evaluate the optimal value of $B_z$ (cf. histograms 3,4 in Fig. 4).

The above definition of short-circuiting assumes that connection of two planes (i.e. virtual electrodes) with an electroconductive filament (which, for Figs. 1-4, is of the length of few tens of basic blocks) is statistically representative for describing the short-circuiting at a much longer distance. Indeed, for a certain initial density of blocks, the dependence of short-circuiting time on the distance, ΔZ, between two virtual planes, connectable with an electroconductive filament, tends, with increasing ΔZ, to a certain limit. This is illustrated with Fig. 5 and Fig. 6.

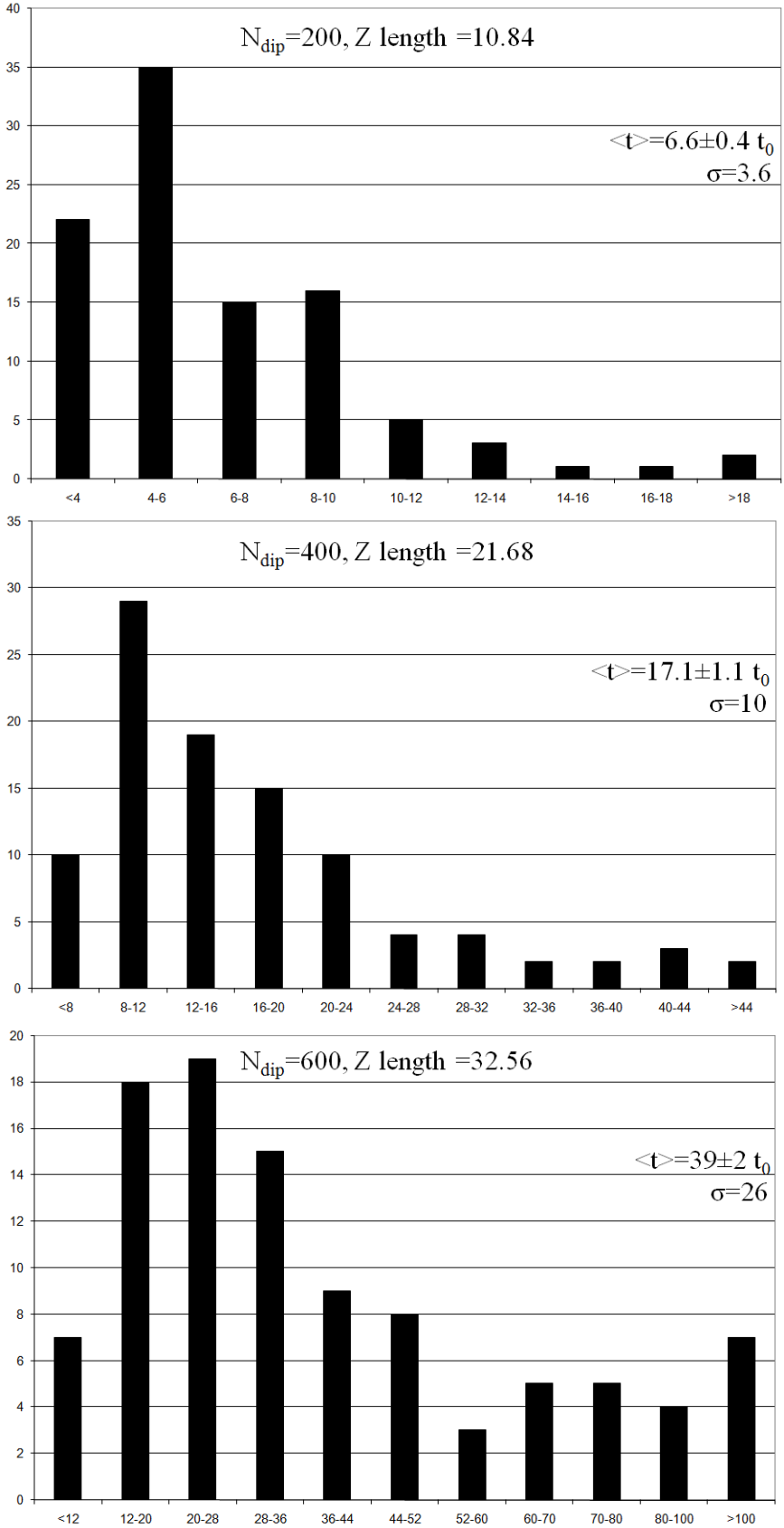

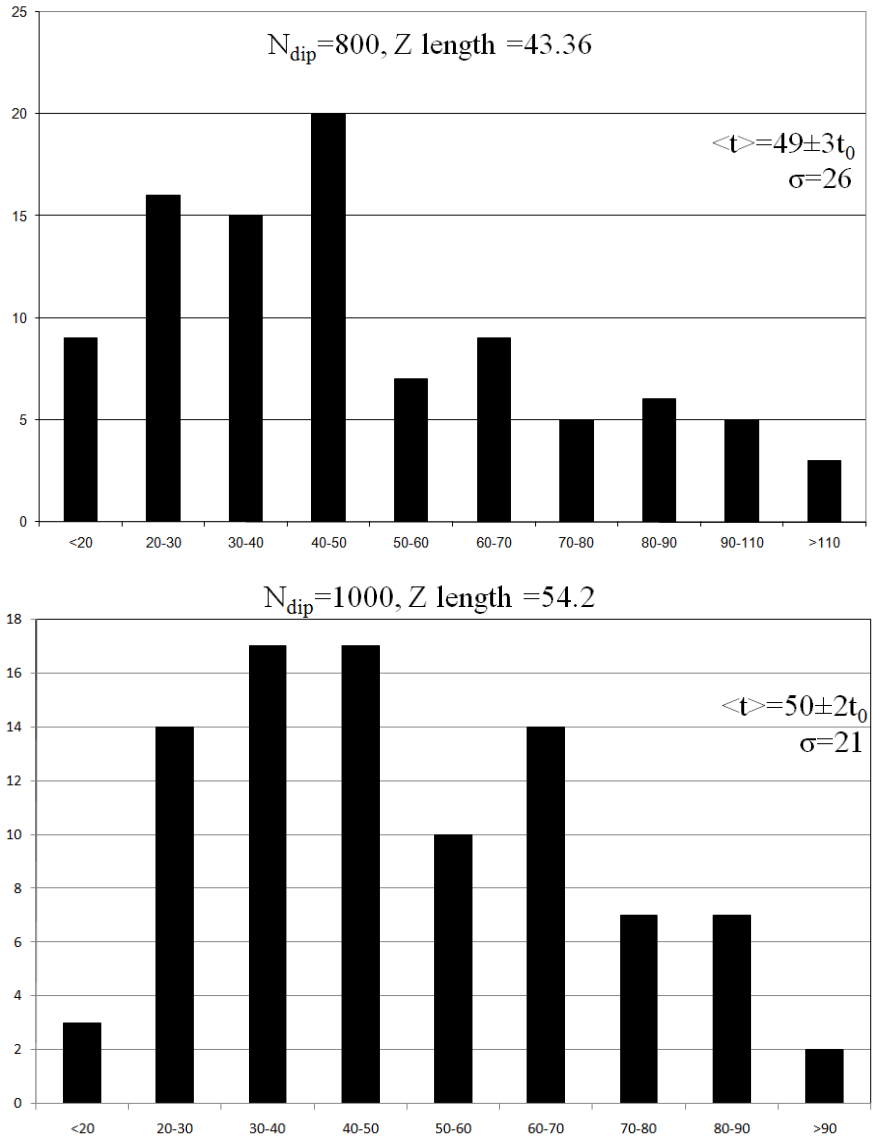

Fig. 5. Statistics of short-circuiting time for 200, 400, 600, 800 and 1000 dipoles (specifically, for various length and identical cross-section of the initial box, and identical initial density of blocks, cf. Fig. 6). Here $B_z=3B_0$ The last pocket gives the sum of distribution's tail.

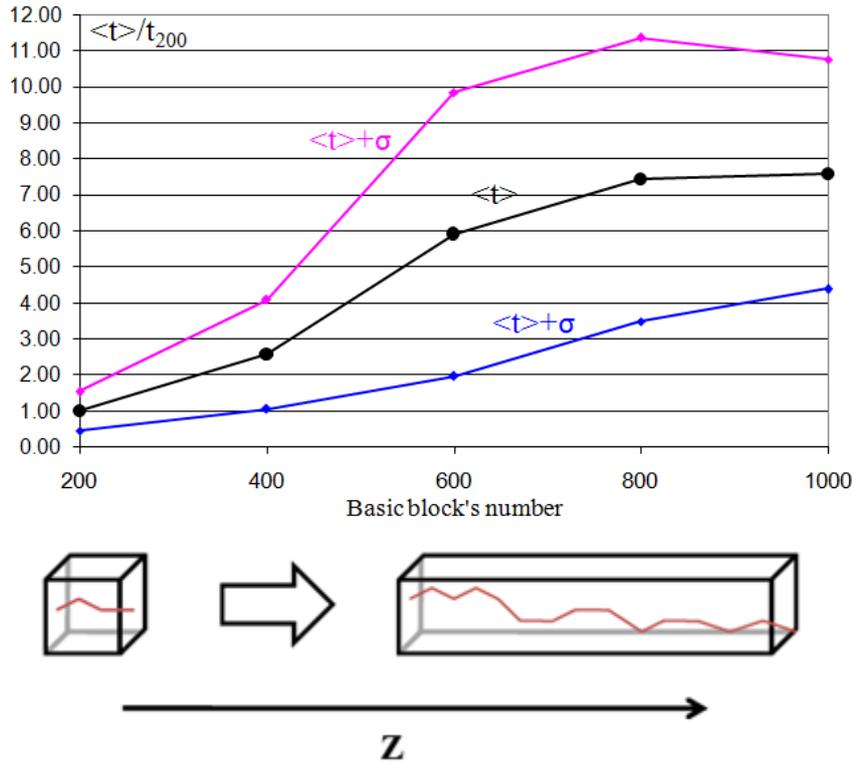

Fig. 6. Dependence of short-circuiting time (relative to that in the case of 200 basic blocks) on the distance between biased electrodes (or, equivalently, on the number of blocks) for identical initial density of blocks and identical transverse size of the spatial region. The central curve gives the average time <t> for 100 computations with random initial distribution in space and in effective magnetic (and electric) charge (similarly to Fig. 4), while upper and lower curves give average time with, respectively, added and subtracted mean-root-square deviation time.

Statistical analysis of short-circuiting time allows tracing the dynamic percolation of electric conductivity and shows a decrease of percolation threshold for volume fraction (vol%), as compared with the observed percolation of carbon nanotubes (CNTs) in liquids and polymer composites. Despite such a comparison is conditional, because of much different conditions of networking of basic blocks, the common main feature of the process is the same, namely formation of a pathway between the electrodes in an electrical circuit. This may be due to a static (statistical) process, when nanotubes are dispersed in an electrically insulating matrix and at a critical concentration the percolation of the conductive particles is attained [32], or due to a dynamic process, when nanotubes are suspended in electrically insulating liquid [33,34].

The latter case with an additional agitation of the liquid (e.g., by ultrasonication) is most close to our case (self-assembling of magnetized electroconductive, electrically charged and electrically self-screened blocks, in a braking medium, non-electric and non-magnetic one). The mobility of CNT blocks in liquid is shown [35] to decrease the percolation threshold as compared with the case of undisturbed liquid (see Fig. 4 in [35]) and with the most reports about solid composites with randomly aligned CNTs (cf., e.g., the most recent survey [36] and Fig. 1 therein). In our computations, the percolation takes place already at the blocks' volume fraction ~ 2÷4·10⁻² vol% (for histograms in Fig. 4) even for rather low values of aspect ratio (length to diameter, L/D ~ 20), for which the static (statistical) percolation threshold is expected to be ~ D/L that is as high as ~ 5 vol%.

## 3. Networking of filaments and a trend towards a fractal skeletal structuring

Here we present the results of a continuous modeling of aggregation a random ensemble of the blocks between the biased electrodes. This includes the short-circuiting stage of evolution and traces the dynamics of pinching of electric current filaments to show the interplay of all the magnetic and electric mechanisms of filaments' networking, including the dynamo effect. The computation is carried out with the parallel code SELFAS-2. It is based on the formalism of the model [1] and extends this model (and respective calculations [2-6]) to allow for
(i) dynamic coupling of filaments to electrodes during short-circuiting (the interaction with the electrodes is modeled by a 1D potential well for magnetic charges on the tips of the blocks),
(ii) electric current dynamics after short-circuiting for a given voltage $U$ and electric resistance $R_0$ of blocks, which is assumed to be determined by the resistance of the tips (the inductance of filaments is ignored).
The interaction of electric current in the blocks at a distance $r$ (similarly to [1], expressed in the units of magnetic attraction force at unit length) is expressed in terms of parameter $F_{0JJ}$ [1]:

$$\vec{F}_{Amper} = \frac{F_{0JJ} I_i I_j \vec{L}_i \times (\vec{L}_j \times \vec{r}_{ij})}{r_{ij}^3}, \qquad F_{0JJ} = \left(\frac{J_0}{J_{ZM}}\right)^2, \qquad (3)$$

$$J_{ZM} = cZ_M e/L = \left(\frac{10nm}{L}\right) Z_M \, 5 \cdot 10^{-3} \, A, \qquad (4)$$

where the current $I_i$ in the i-th block is expressed in the units of the current $J_0$ and is defined by the electric circuit equations and the parameter

$$\delta_{J0} = \frac{U L}{R_0 J_0 \Delta Z}, \qquad (5)$$

where $\Delta Z$ is the distance between electrodes.

The results of modeling are presented in Figs 7-16.

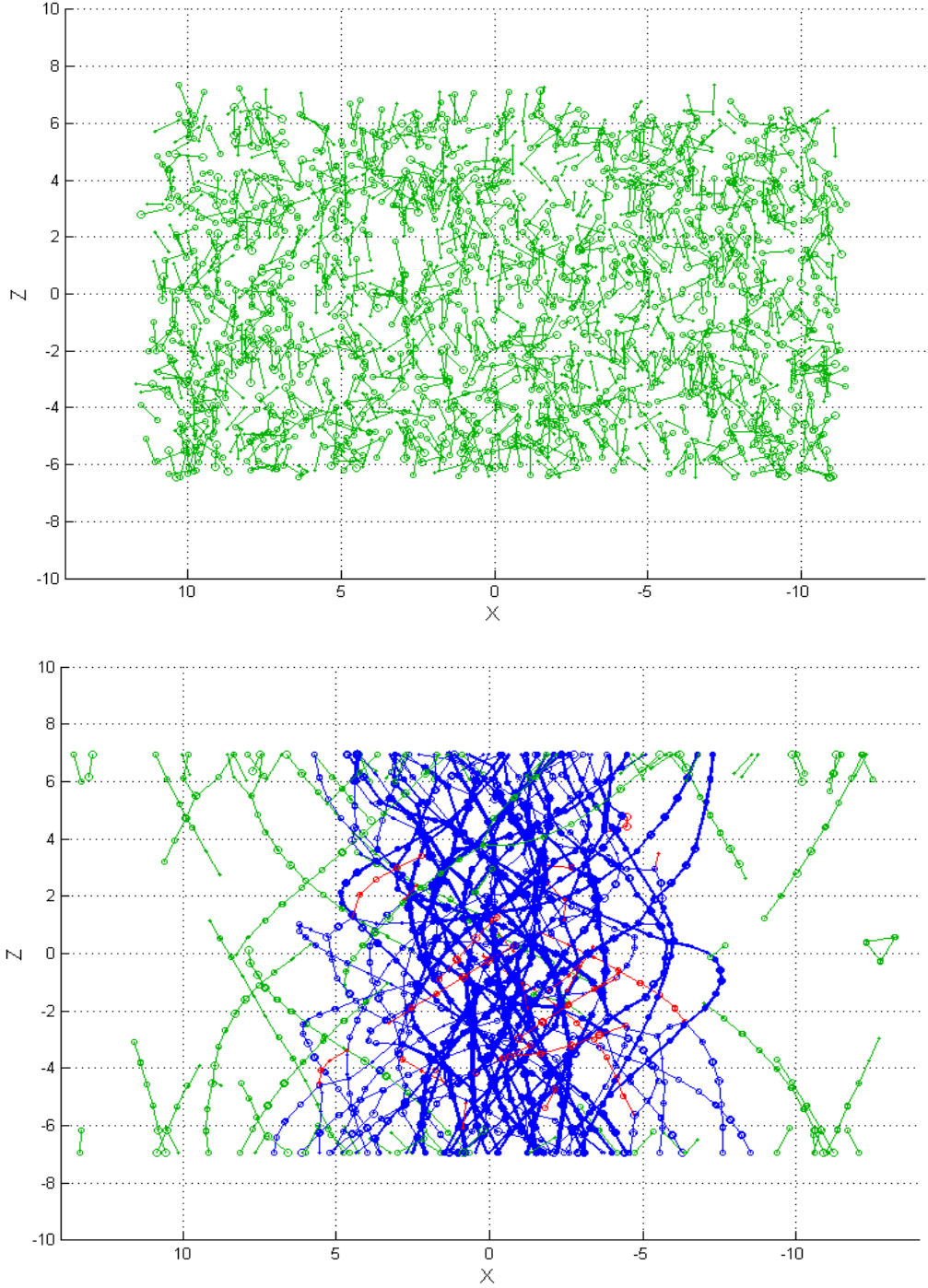

**Fig. 7.** Side-on view of the ensemble of basic blocks at time t=0 and t=300 (in units of $t_0$, Eq. (1)). Short-circuited filaments are shown with blue color, while the dead ends within the network of short-circuited filaments are shown with red. Block's number $N_{dip}$ = 1200, block's inverse aspect ratio D/L = 0.06, initial space dimensions = [(-11.1,11.1);(-11.1,11.1);(-6.45,6.45)] (in units of block's length *L*), electric screening radius $r_D$ = 1.00, $F_{0JJ}$ = 0.250 (Eq. (3)), brake coefficients for tip-tip collision, $K_{brake}$ = 100, and for brake in a ambient medium, $M_{brake}$ = 1.5, $\delta_{J0}$ = 2.86 (Eq. (5)), external magnetic field $B_{ext}$ = [0,0,1.5] (in units of $B_0$, Eq. (2)), radius of (z-directed, with center at x=y=0) electric current filament $R_{plas}$ = 6, total longitudinal electric current through plasma filament $J_{zPlas}$=2.5 $J_{ZM}$.

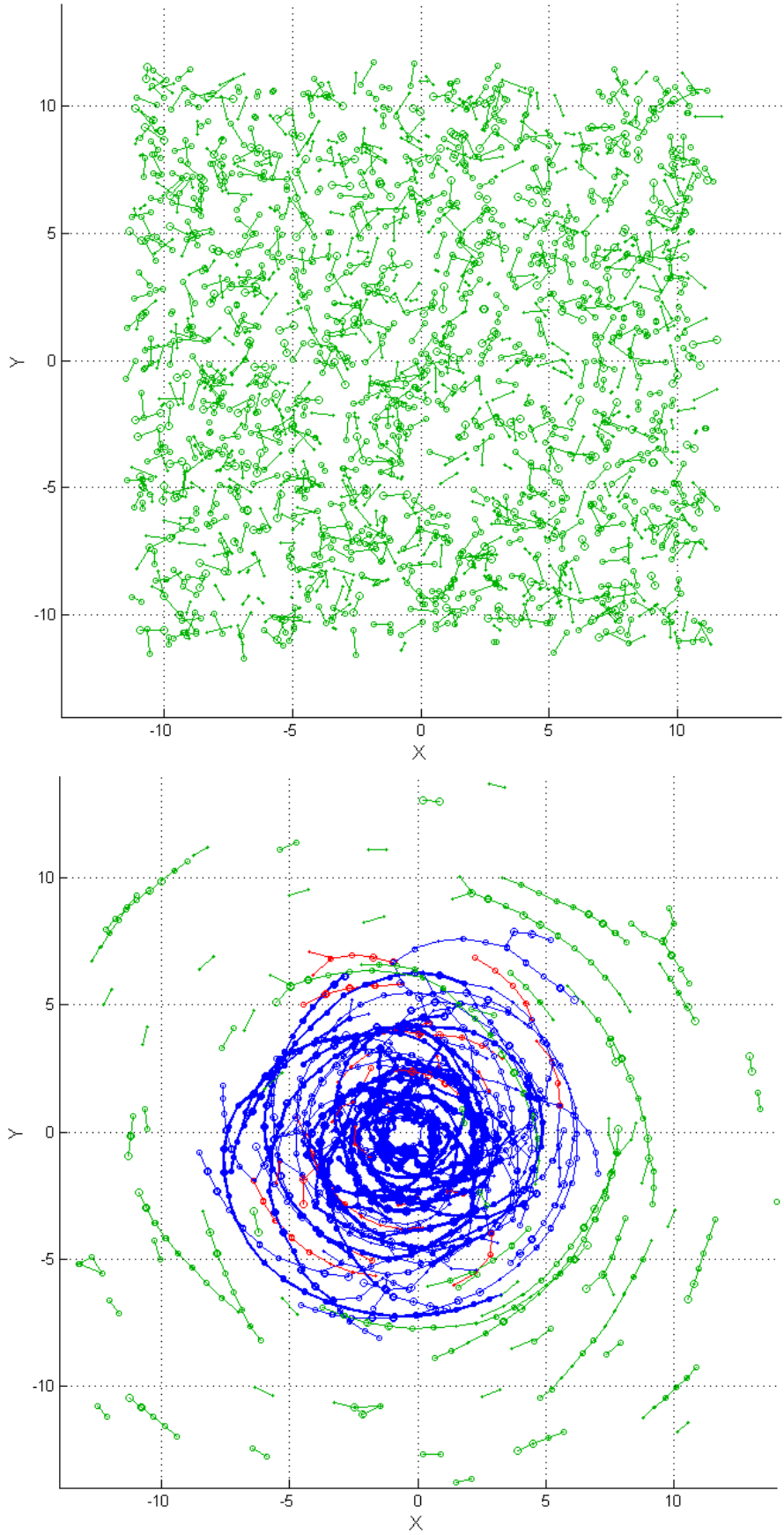

**Fig. 8.** Top view of the ensemble of basic blocks at time t=0 and t=300 $t_0$ for the conditions of Fig. 7.

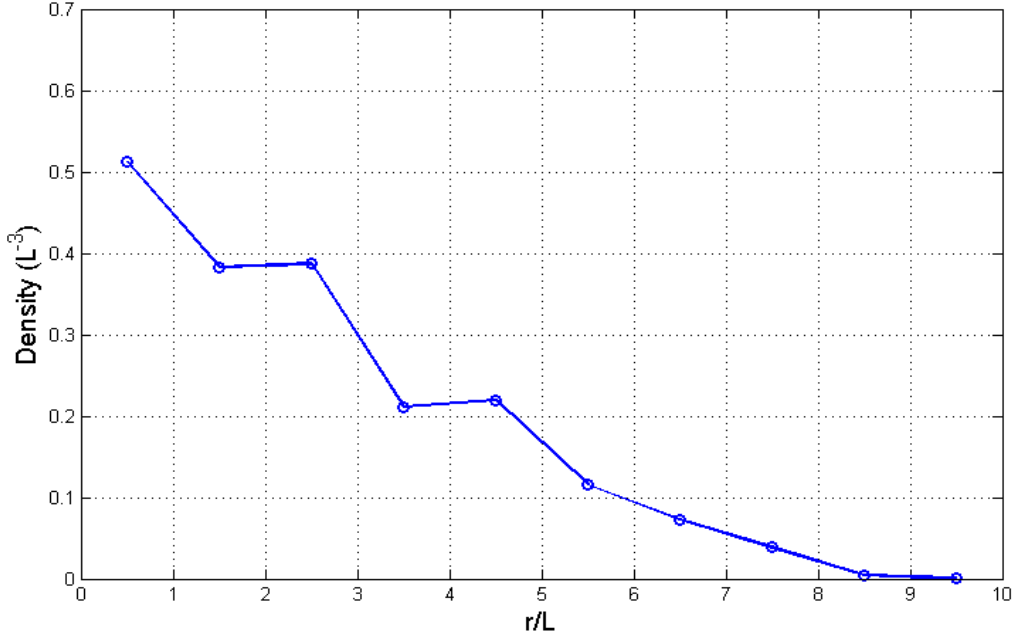

**Fig. 9.** Radial profile of density of basic blocks (averaged over longitudinal direction and azimuthal angle) at time t=300 $t_0$ for the conditions of Fig. 7.

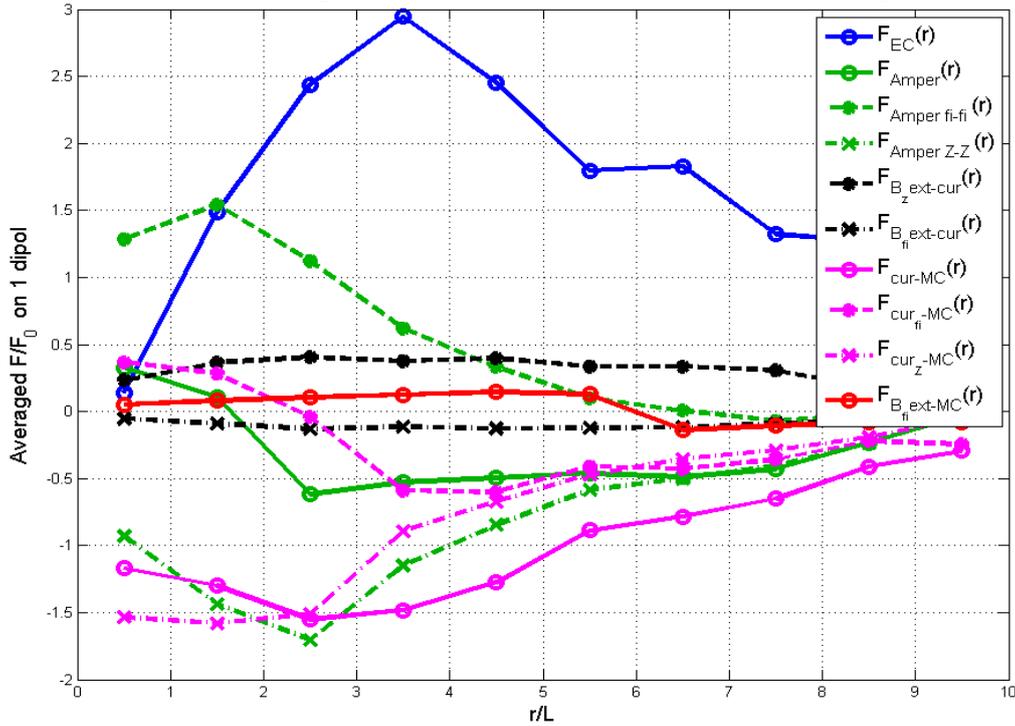

**Fig. 10.** Radial profile of the forces (averaged over longitudinal direction and azimuthal angle) acting at a basic block at time t=300 $t_0$ for the conditions of Fig. 7. The following forces are shown: EC – electric repulsion, Amper – interaction of electric currents through filaments composed of basic blocks, "Amper fi-fi" ("Amper Z-Z") – interaction of azimuthal (longitudinal) components of these electric currents, "$B_z$ext-cur" ("$B_{fi}$ext-cur") – interaction of these current with external longitudinal (azimuthal) magnetic field, "cur-MC" ("$cur_{fi}$-MC", "$cur_Z$-MC") – interaction of magnetic dipole with magnetic field of electric current (its azimuthal, longitudinal components) through the filaments composed of basic blocks, "$B_{fi}$ext-MC" – interaction of magnetic dipole with azimuthal magnetic field of plasma filament.

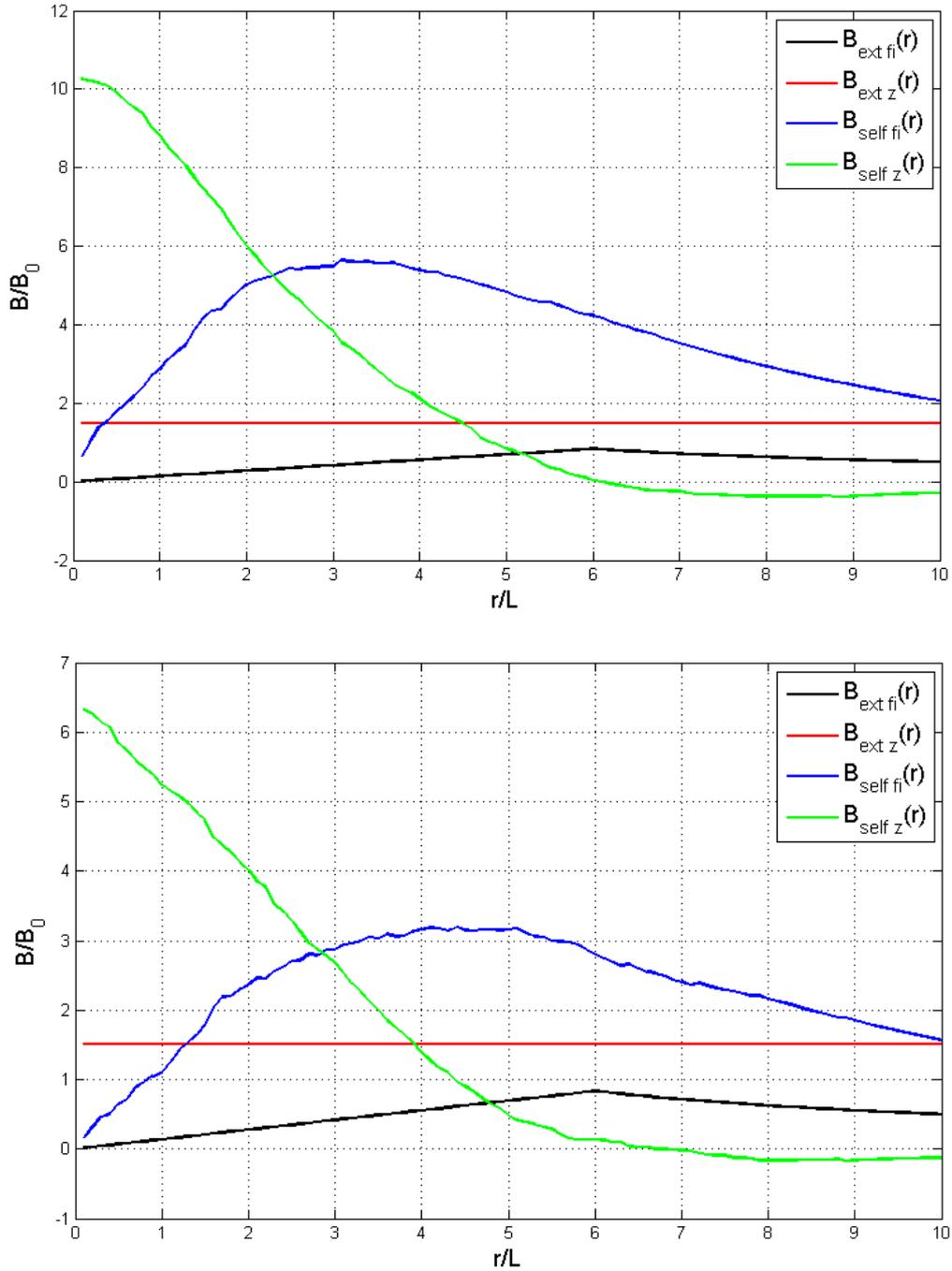

**Fig. 11.** Radial profile of magnetic fields, averaged over azimuthal angle, at time t=300 $t_0$ in the center of the column, z=0, (upper figure) and in the edge, z=6, (lower figure) for the conditions of Fig. 7. The following fields are shown: "ext fi" – azimuthal magnetic field of z-directed plasma filament, "ext z" - external z-directed magnetic field, "self fi" - azimuthal magnetic field of electric current through filaments composed of basic blocks, "self z" – z component of these currents.

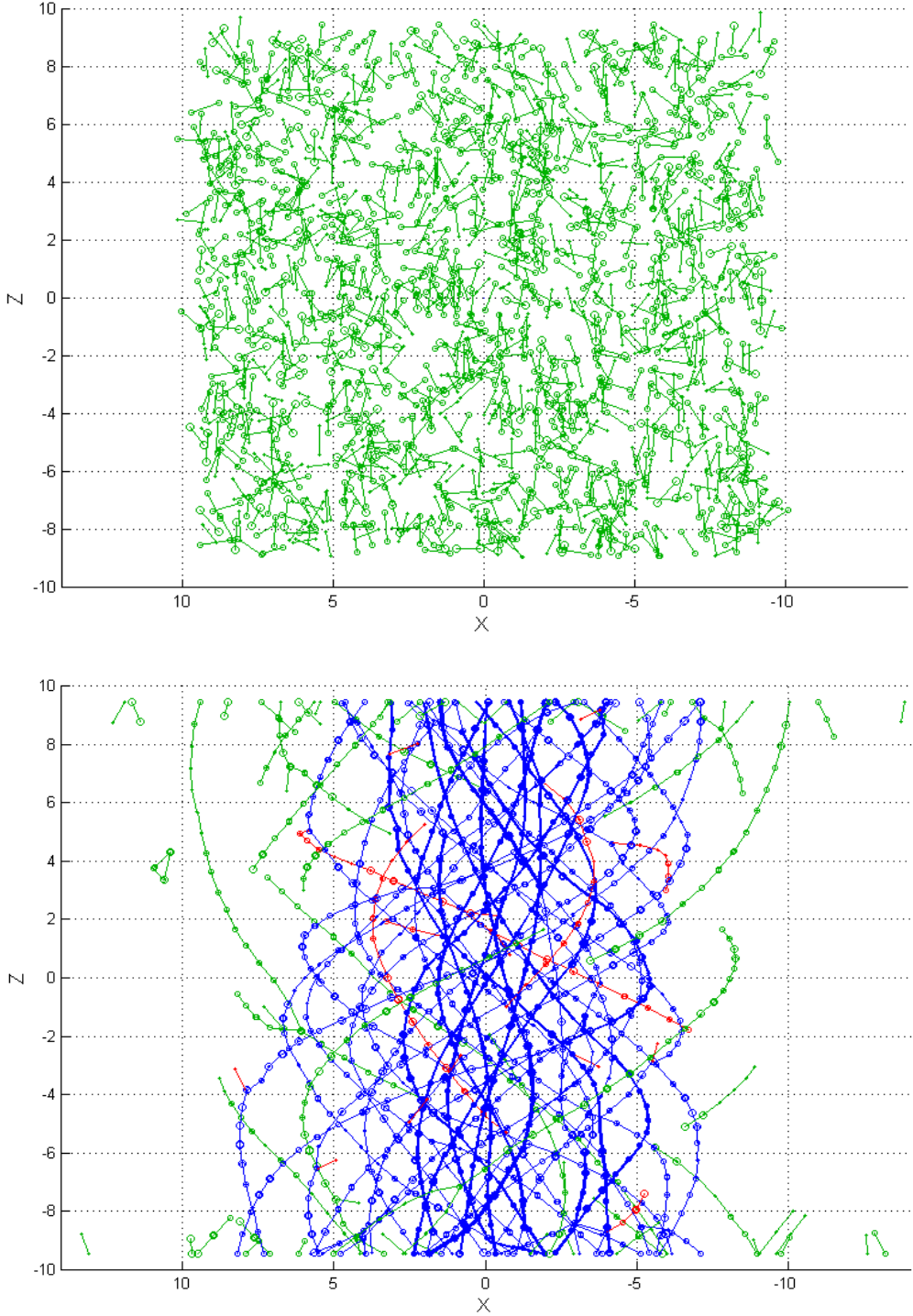

**Fig. 12.** Side-on view of the ensemble of basic blocks at time t=0 and t=300 (in units of $t_0$, Eq. (1)). Short-circuited filaments are shown with blue color, while the dead ends within the network of short-circuited filaments are shown with red. Block's number Ndip = 1200, block's inverse aspect ratio D/L = 0.06, initial space dimensions = [(-9.2,9.2);(-9.2,9.2);(-9.45,9.45)] (in units of block's length *L*), electric screening radius $r_D$ = 1.00, $F_{0JJ}$ = 0.250 (Eq. (3)), brake coefficients for tip-tip collision, $K_{brake}$ = 100, and for brake in a ambient medium, $M_{brake}$ = 1.5, $\delta_{J0}$ = 2.1 (Eq. (5)), external magnetic field $B_{ext}$ = [0,0,1.5] (in units of $B_0$, Eq. (2)), radius of (z-directed, with center at x=y=0) electric current filament $R_{plas}$ = 6, total longitudinal electric current through plasma filament $J_{zPlas}$=7.5 $J_{ZM}$.

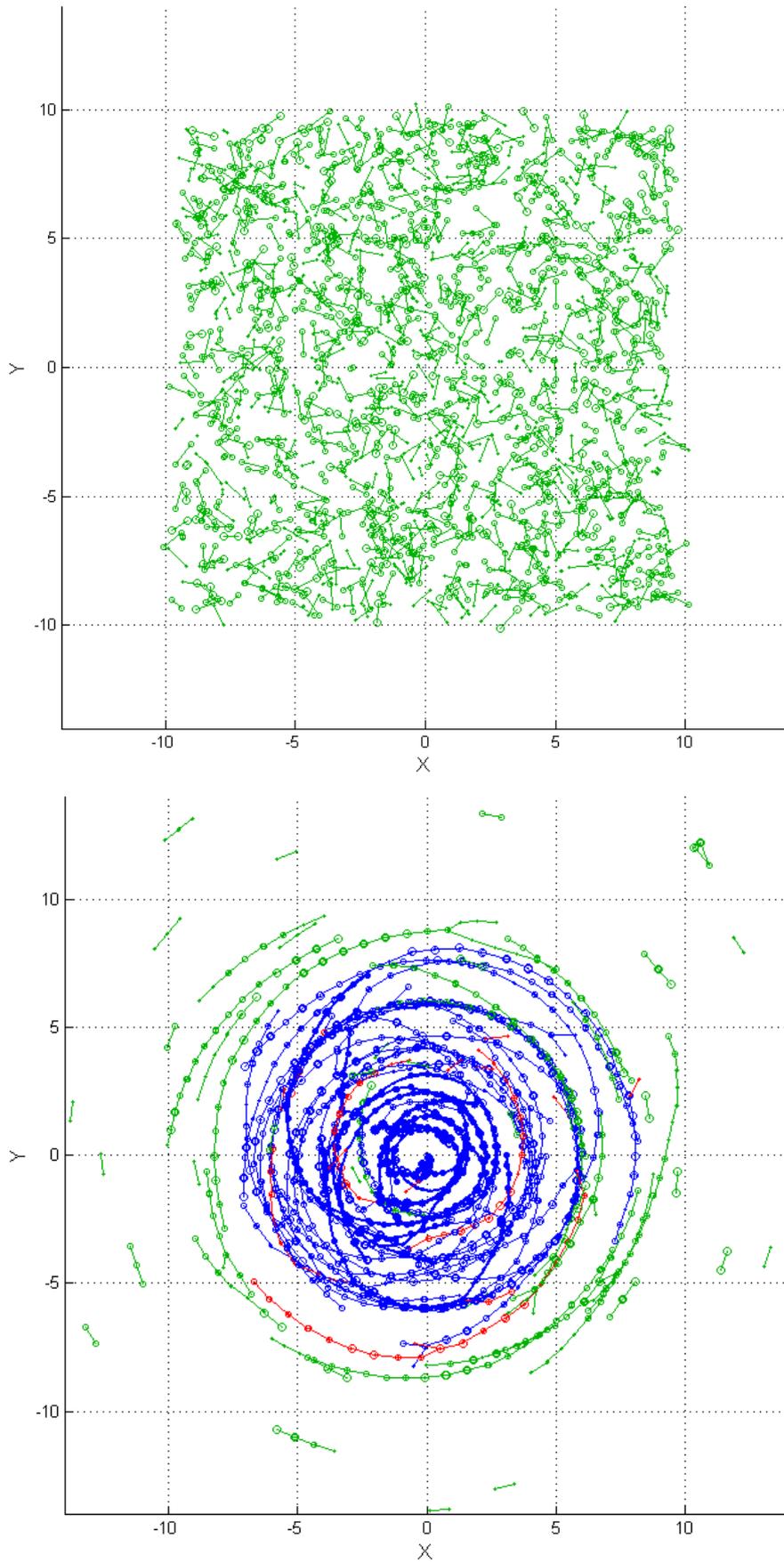

**Fig. 13.** Top view of the ensemble of basic blocks at time t=0 and t=300 $t_0$ for the conditions of Fig. 12.

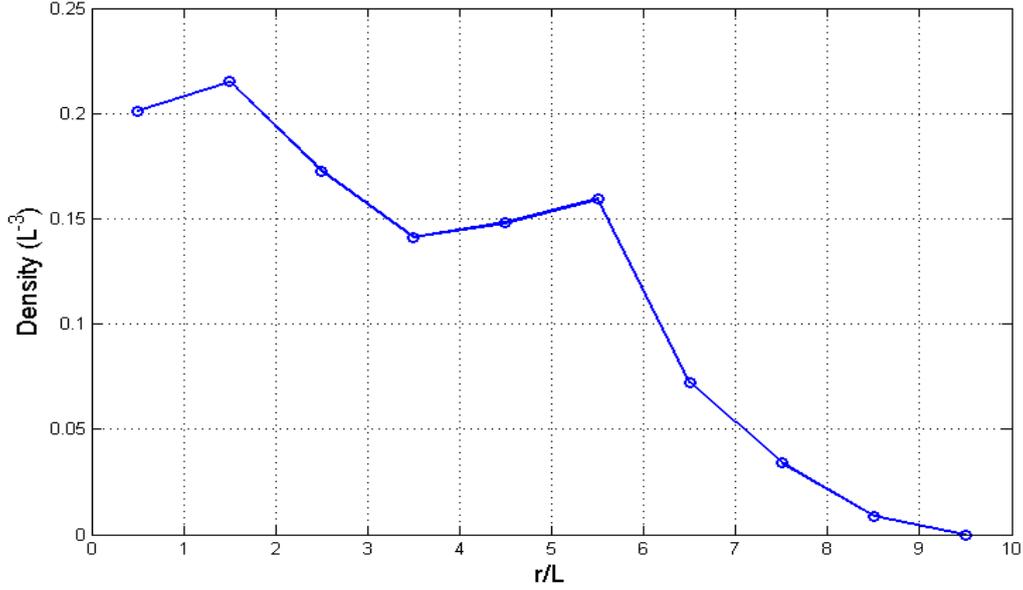

**Fig. 14.** Radial profile of density of basic blocks (averaged over longitudinal direction and azimuthal angle) at time $t=300\ t_0$ for the conditions of Fig. 12.

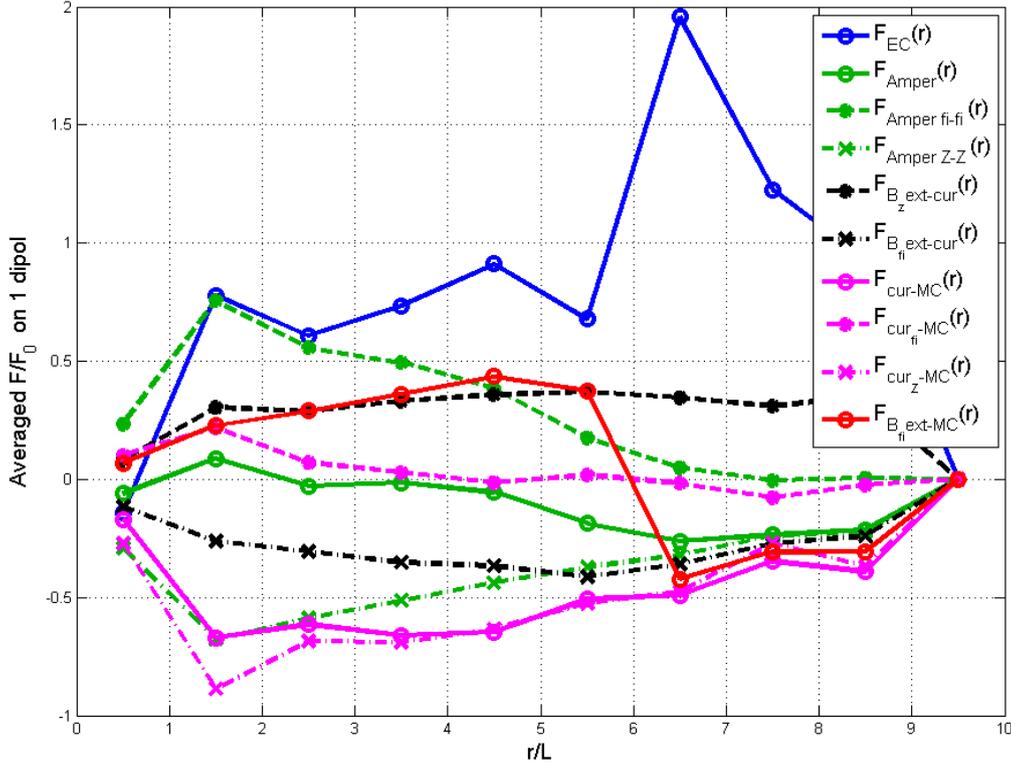

**Fig. 15.** Radial profile of the force (averaged over longitudinal direction and azimuthal angle) acting at a basic block at time $t=300\ t_0$ for the conditions of Fig. 12. The following forces are shown: EC – electric repulsion, Amper – interaction of electric currents through filaments composed of basic blocks, "Amper fi-fi" ("Amper Z-Z") – interaction of azimuthal (longitudinal) components of these electric currents, "$B_z$ext-cur" ("$B_{fi}$ext-cur") – interaction of these current with external longitudinal (azimuthal) magnetic field, "cur-MC" ("cur$_{fi}$-MC", "cur$_Z$-MC") – interaction of magnetic dipole with magnetic field of electric current (its azimuthal, longitudinal components) through the filaments composed of basic blocks, "$B_{fi}$ext-MC" – interaction of magnetic dipole with azimuthal magnetic field of z-directed plasma filament.

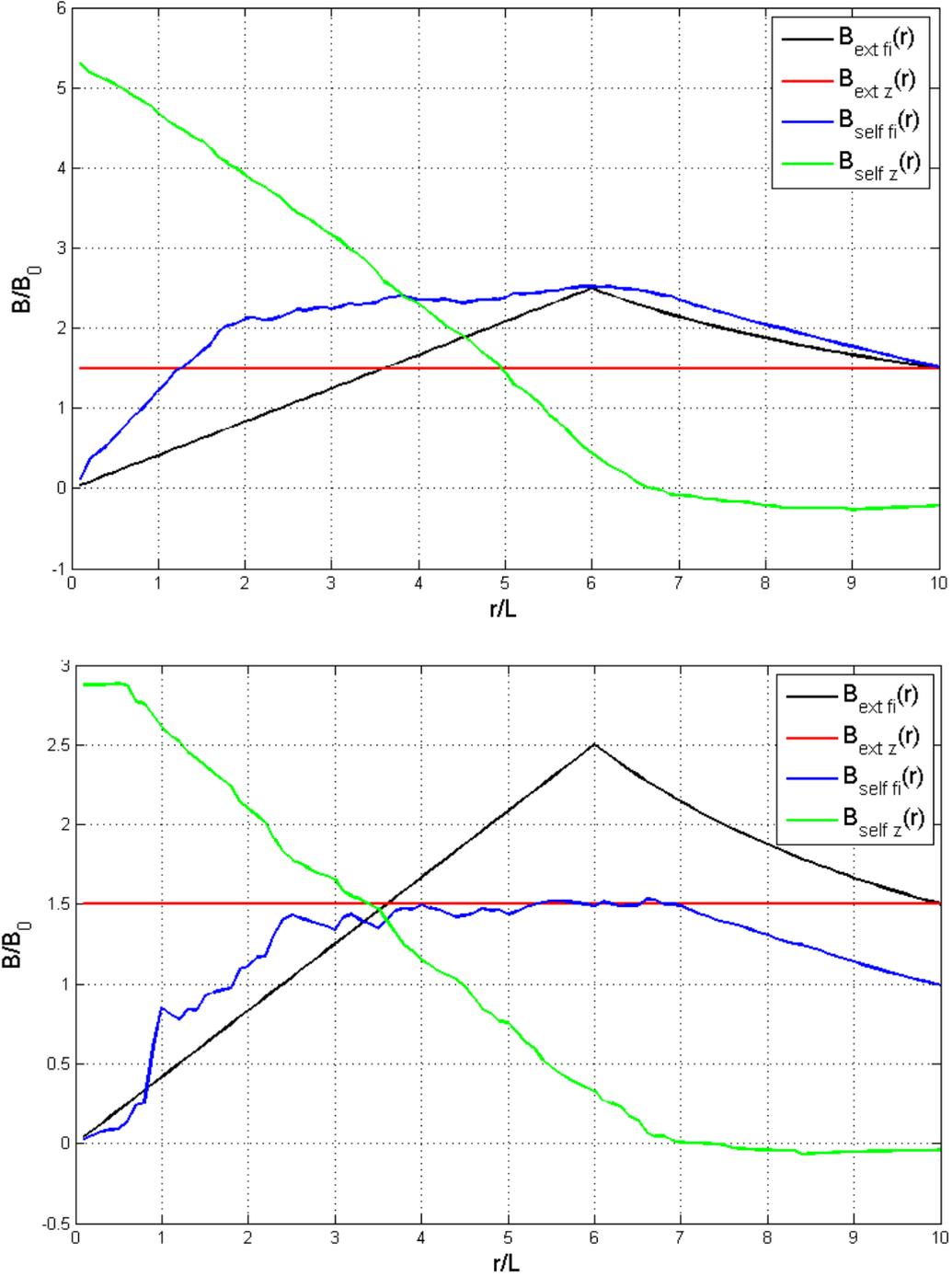

**Fig. 16.** Radial profile of magnetic fields (averaged over azimuthal angle) at time t=300 $t_0$ in the center of the column, z=0, (upper figure) and in the edge, z=9, (lower figure) for the conditions of Fig. 12. The following fields are shown: "ext fi" – azimuthal magnetic field of z-directed plasma filament, "ext z" - external z-directed magnetic field, "self fi" - azimuthal magnetic field of electric current through filaments composed of basic blocks, "self z" – z component of these currents.

The figures 11 and 16 gives an evidence for generation of longitudinal magnetic field and, respectively, of a bigger magnetic dipole. This extends previous evidence for such a dynamo effect found in simulation of a bunch of initially linear filaments "artificially" composed of basic blocks (see Fig. 5 in [5]) to the case of modeling starting from a random ensemble of basic blocks. It follows that a trend towards a fractal skeletal structuring (namely, repeat of original basic block at a larger length scale) may be in a much wider class of initial conditions.

**4. Conclusions**

The self-assembling of quasi-linear filaments from an ensemble of randomly situated basic blocks and the electric short-circuiting between biased electrodes are shown to be supported by the alignment of blocks in an external magnetic field. Statistical analysis of short-circuiting time allows tracing the dynamic percolation of electric conductivity. These results suggest the positive role of external magnetic field for electrodynamic self-assembling of macroscopic skeletal structures, identified in a broad range of length scales in laboratory electric discharges and space [11-13], from the blocks with above-mentioned electrodynamic properties (presumably, carbon nanotubes and similar nanostructures). This positive impact seems to be especially strong in the electric discharges in a strong toroidal magnetic field.

A continuous numerical modeling of various stages of aggregation in a random ensemble of the blocks between the biased electrodes is carried out with a parallel numerical code. This allows tracing the dynamics of pinching of the emerged electric current filaments and the interplay of all the magnetic and electric mechanisms of filaments' networking, including the dynamo effect. A trend towards a fractal skeletal structuring (namely, repeat of original basic block at a larger length scale) is illustrated with the evidence for generation of a bigger magnetic dipole composed of basic blocks.

**Acknowledgments**

This work is supported by the Russian Foundation for Basic Research (project RFBR 05-08-65507) and the European project EGEE-III (Enabling Grids for E-sciencE). The distributed computations are carried out in the GRID system RDIG (Russian Data Intensive Grid, virtual organization RFUSION). The parallel computations are carried out in the HPC cluster of the RRC "Kurchatov Institute".